\documentclass[aps,prl,twocolumn,superscriptaddress,showpacs]{revtex4}
\usepackage{graphicx}
\usepackage{helvet}

\begin{document}

\title{Mott Transition from a Spin Liquid to a Fermi Liquid in the Spin-Frustrated Organic Conductor $\kappa$-(ET)$_2$Cu$_2$(CN)$_3$}

\author{Y. Kurosaki}
\affiliation{Department of Applied Physics, University of Tokyo, Bunkyo-ku, Tokyo, 113-8656, Japan.}
\author{Y. Shimizu}
\altaffiliation[Present address:]{RIKEN, Wako, Saitama 351-0198, Japan.}
\affiliation{Department of Applied Physics, University of Tokyo, Bunkyo-ku, Tokyo, 113-8656, Japan.}
\affiliation{Division of Chemistry, Kyoto University, Sakyo-ku, Kyoto, 606-8502, Japan.}
\author{K. Miyagawa}
\affiliation{Department of Applied Physics, University of Tokyo, Bunkyo-ku, Tokyo, 113-8656, Japan.}
\affiliation{CREST, Japan Science and Technology Agency, Kawaguchi, Saitama 332-0012, Japan.}
\author{K. Kanoda}
\affiliation{Department of Applied Physics, University of Tokyo, Bunkyo-ku, Tokyo, 113-8656, Japan.}
\affiliation{CREST, Japan Science and Technology Agency, Kawaguchi, Saitama 332-0012, Japan.}
\author{G. Saito}
\affiliation{Division of Chemistry, Kyoto University, Sakyo-ku, Kyoto, 606-8502, Japan.}

\date{\today}

\begin{abstract}
	Pressure-temperature phase diagram of the organic Mott insulator $\kappa$-(ET)$_2$Cu$_2$(CN)$_3$, a model system of the spin liquid on triangular lattice, has been investigated by $^1$H NMR and resistivity measurements. 
	The spin-liquid phase is persistent before the Mott transition to the metal or superconducting phase under pressure. 
	At the Mott transition, the spin fluctuations are rapidly suppressed and the Fermi-liquid features are observed in the temperature dependence of the spin-lattice relaxation rate and resistivity. 
	The characteristic curvature of Mott boundary in the phase diagram highlights a crucial effect of the spin frustration on the Mott transition. 
\end{abstract}



\maketitle

	Magnetic interaction on the verge of Mott transition is one of the chief subjects in the physics of strongly correlated electrons, because striking phenomena such as unconventional superconductivity emerge from the mother Mott insulator with antiferromagnetic (AFM) order. 
	Examples are transition metal oxides such as V$_2$O$_3$ and La$_2$CuO$_4$, in which localized paramagnetic spins undergo the AFM transition at low temperatures \cite{Ref1}. 
	The ground state of Mott insulator is, however, no more trivial when the spin frustration works between the localized spins. 
	Realization of spin liquid has attracted much attention since a proposal of the possibility in triangular-lattice Heisenberg antiferromagnet \cite{Ref2}.
	Owing to the extensive materials research, some examples of the possible spin liquid have been found in systems with triangular and kagom\'e lattices, such as the solid $^3$He layer \cite{Ref3}, Cs$_2$CuCl$_4$ \cite{Ref4} and $\kappa$-(ET)$_2$Cu$_2$(CN)$_3$ \cite{Ref5}.
	Mott transitions between metallic and insulating spin-liquid phases are an interesting new area of research. 

	The layered organic conductor $\kappa$-(ET)$_2$Cu$_2$(CN)$_3$ is the only spin-liquid system to exhibit the Mott transition, to the authors' knowledge \cite{Ref5}. 
	The conduction layer in $\kappa$-(ET)$_2$Cu$_2$(CN)$_3$ consists of strongly dimerized ET [bis(ethlylenedithio)-tetrathiafulvalene] molecules with one hole per a dimer site, so that the on-site Coulomb repulsion inhibits the hole transfer \cite{Ref6}. 
	In fact, it is a Mott insulator at ambient pressure and becomes a metal/superconductor under pressure \cite{Ref7}. 
	Taking the dimer as a unit, the network of inter-dimer transfer integrals forms a nearly isotropic triangular lattice, and therefore the system can be modeled to a half-filled band system with strong spin frustration on the triangular lattice. 
	At ambient pressure, the magnetic susceptibility behaved as the triangular-lattice Heisenberg model with an AFM interaction energy $J$ $\sim$ 250 K \cite{Ref5, Ref8}. 
	Moreover, the $^1$H NMR measurements provided no indication of long-range magnetic order down to 32 mK. 
	These results suggested the spin liquid state at ambient pressure. 
	Then the Mott transition in $\kappa$-(ET)$_2$Cu$_2$(CN)$_3$ under pressure may be the unprecedented one without symmetry breaking, if the magnetic order does not emerge under pressure up to the Mott boundary. 

	\begin{figure}
	\includegraphics[scale=0.38]{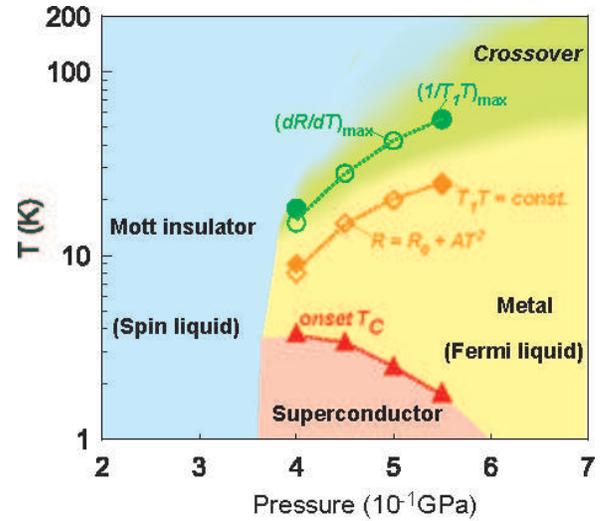}
	\caption{\label{Fig1}The pressure-temperature phase diagram of $\kappa$-(ET)$_2$Cu$_2$(CN)$_3$, constructed on the basis of the resistance and NMR measurements under hydrostatic pressures. 
	The Mott transition or crossover lines were identified as the temperature where $1/T_1T$ and $dR/dT$ show the maximum as described in the text. 
	The upper limit of the Fermi liquid region was defined by the temperatures where $1/T_1T$ and $R$ deviate from the Korringa's relation and $R_0 + AT^2$, respectively. 
	The onset superconducting transition temperature was determined from the in-plane resistance measurements.
	}
	\end{figure}
	In this Letter, we report on the NMR and resistance studies of the Mott transition in $\kappa$-(ET)$_2$Cu$_2$(CN)$_3$ under pressure. 
	The result is summarized by the pressure-temperature ($P-T$) phase diagram in Fig.1. 
	The Mott insulating phase is revealed to be the spin-liquid state without magnetic order. 
	The Mott transition/crossover and the entrance into Fermi liquid regime are identified both from the characteristic nuclear spin-lattice relaxation rate ($1/T_1$) and resistance behaviors. 
	The profile of the Mott transition boundary in the $P-T$ diagram is contrast to the conventional Mott insulator with AFM order, reflecting that the large spin entropy is remained even at low temperatures in the frustrated spin-liquid state. 

	The samples were prepared by the standard electrochemical method \cite{Ref7, Ref9}. 
	The hydrostatic pressure was applied by using the cramp-type BeCu cell with the proton-free DEMNUM S-20 oil as the pressure medium. 
	The pressures quoted in this Letter are the values monitored at room temperature, but considered to be reduced by about 0.2 GPa at low temperatures due to solidification of the oil. 
	The $^1$H NMR spectra and $1/T_1$ were measured for the polycrystalline sample at 3.6 T by using the solid-echo pulse sequence. 
	The in-plane resistance was measured with the four-probe method at the same pressures applied for the NMR experiments. 

	\begin{figure}
	\includegraphics[scale=0.38]{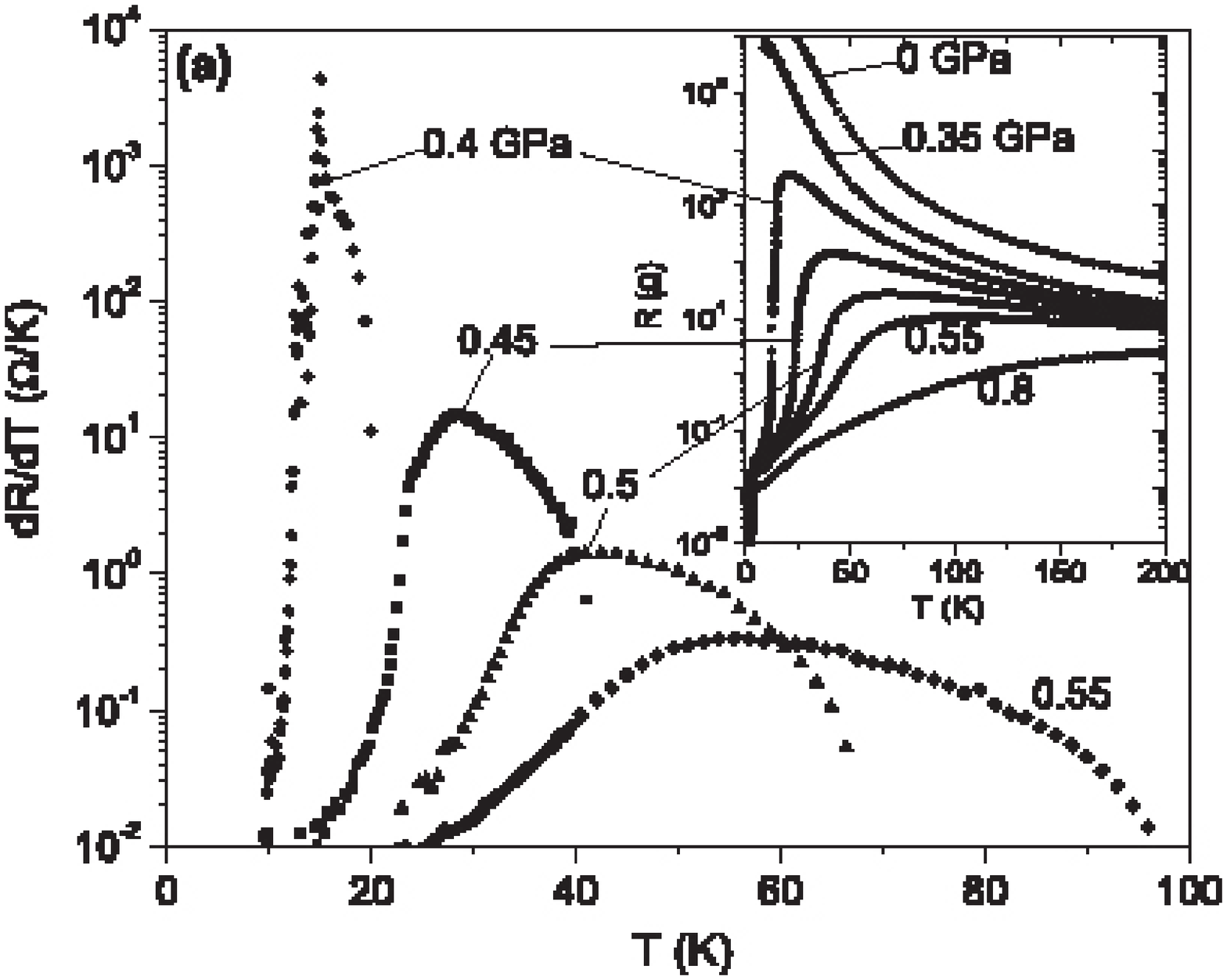}
	\includegraphics[scale=0.38]{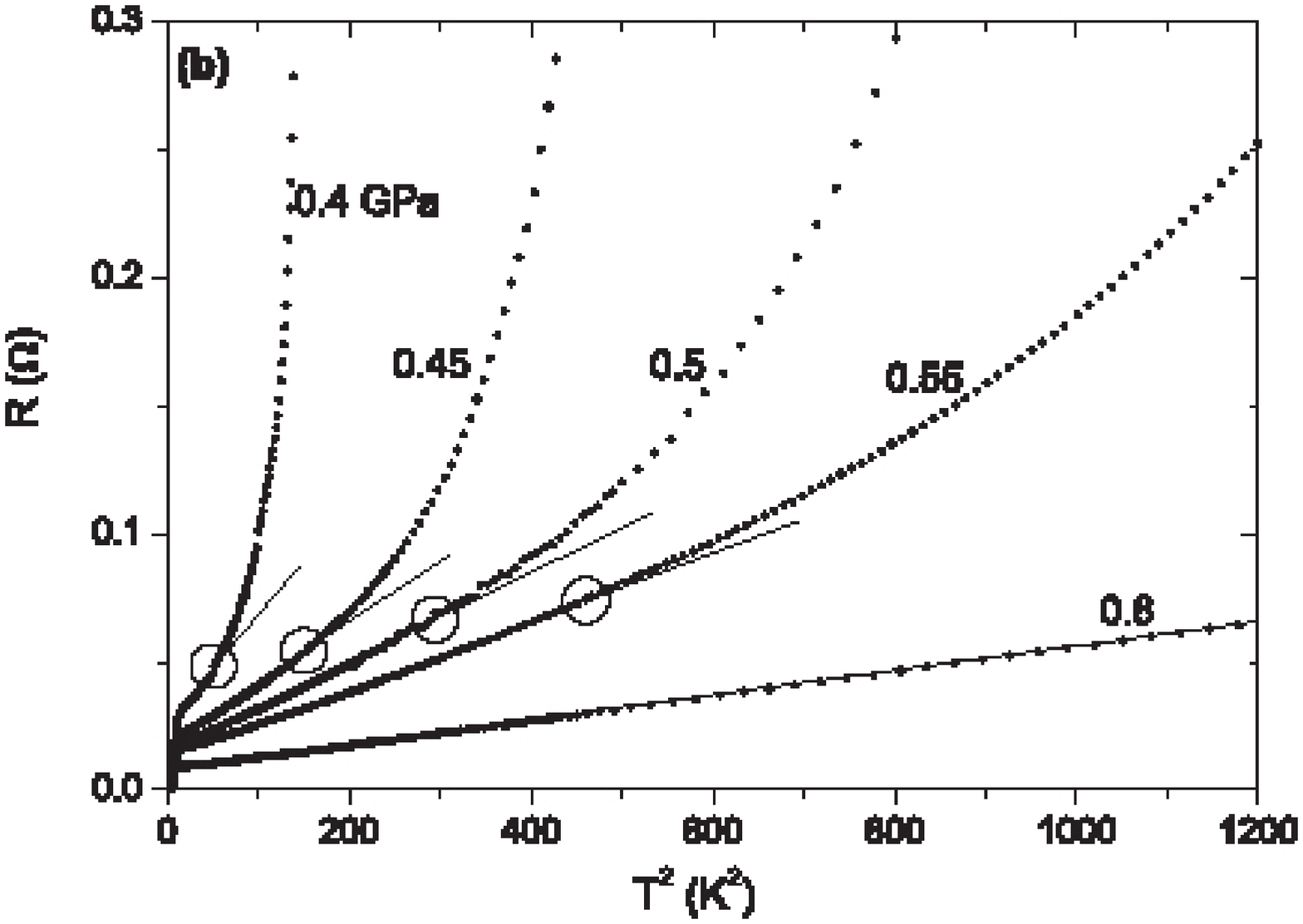}
	\caption{\label{Fig2} 
	(a) Temperature differential of the in-plane resistance of $\kappa$-(ET)$_2$Cu$_2$(CN)$_3$ at various hydrostatic pressures and a zero magnetic field. 
	The inset shows temperature dependence of the resistance. 
	(b) The $T^2$ dependence of the resistance of $\kappa$-(ET)$_2$Cu$_2$(CN)$_3$ under hydrostatic pressures. 
	The circles mark the points where the resistance deviates from the equation $R$ = $R_0$ + $AT^2$. 
	}
\end{figure}
	The result of our resistance measurement is shown in the inset of Fig.2(a). 
	The observed systematic change from insulator to metal is consistent with the previous report \cite{Ref7}. 
	Under a pressure of 0.4 GPa, a drastic insulator-metal transition with a resistance drop by 4 orders of magnitude is observed at 16 K, resulting in a singular peak in the temperature derivative of resistance, $dR/dT$, as displayed in the main panel of Fig.2(a). 
	The temperature giving the maximum in $dR/dT$, where the resistance changes most sharply, is regarded as the Mott transition or crossover, and plotted in Fig.1. 
	The drastic transition at 0.4 GPa suggests the first-order Mott transition as encountered in $\kappa$-(ET)$_2$Cu[N(CN)$_2$]Cl \cite{Ref10, Ref11, Ref12}. 
	At higher pressure, the peak of $dR/dT$ becomes broad and shifts to higher temperatures, indicating that the Mott transition enters into the crossover regime. 
	Figure 2(b) shows the low-temperature resistance in the metallic state as a function of squared temperature. 
	The resistance follows the quadratic temperature dependence, $R$ = $R_0$ + $AT^2$, as expected in a normal Fermi liquid, at low temperatures above the onset temperature of superconducting transition, $T_{\rm C}$. 
	At 0.4 GPa, the Fermi liquid region is quite narrow, say below 8 K, but widens under higher pressures. 
	At the same time, the coefficient $A$ becomes rapidly suppressed. 
	The temperature where the resistance begins to deviate from the $T^2$ law [marked by the circle in Fig.2(b)] is regarded as a measure of the onset into the bad metal where the Fermi-liquid coherence is disturbed, but it is well below the Mott transition and crossover temperature. 
	In fact, these values of the resistance around this region, 0.05-0.1 $\Omega$, which corresponds 0.5-1$\times$10$^{-3}$ $\Omega$ cm in resistivity, is of the order of the Mott-Ioffe-Regel limit \cite{Ref13}, $\rho_0 = \hbar a/e^2$ $\sim$ 10$^{-3}$ $\Omega$ cm, where $a$ is the lattice constant ($\sim$ 10 \AA). 
	To investigate the magnetism and spin excitations around the Mott transition, we conduct NMR measurements at the same pressure points as the resistance measurements. 
	In Fig.3, we show the $^1$H NMR spectra at 1.4 K and various hydrostatic pressures. 
	The linewidth comes from the typical dipole fields between the adjacent $^1$H nuclear spins. 
	We observed neither broadening nor splitting in the NMR spectra at the pressures investigated. 
	This result indicates that the Mott insulating phase has no long-range magnetic order in the whole pressure range up to the critical pressure where the metallic or superconducting phase appears. 
	Namely, the spin-liquid state persists toward the Mott transition. 
	This is consistent with the ground-state calculation of triangular-lattice Hubbard model, which predicts the spin-liquid phase near the Mott transition \cite{Ref14}.
	\begin{figure}
	\includegraphics[scale=0.3]{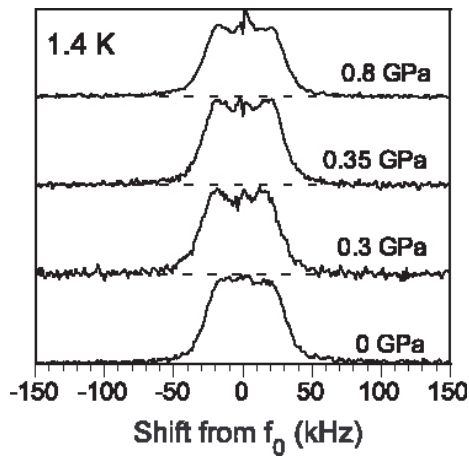}
	\caption{\label{Fig3}
	$^1$H NMR spectra of $\kappa$-(ET)$_2$Cu$_2$(CN)$_3$ at 1.4 K under various hydrostatic pressures. 
	The observed resonance frequency $f_0$ was 156 MHz.}
	\end{figure}

	\begin{figure}
	\includegraphics[scale=0.4]{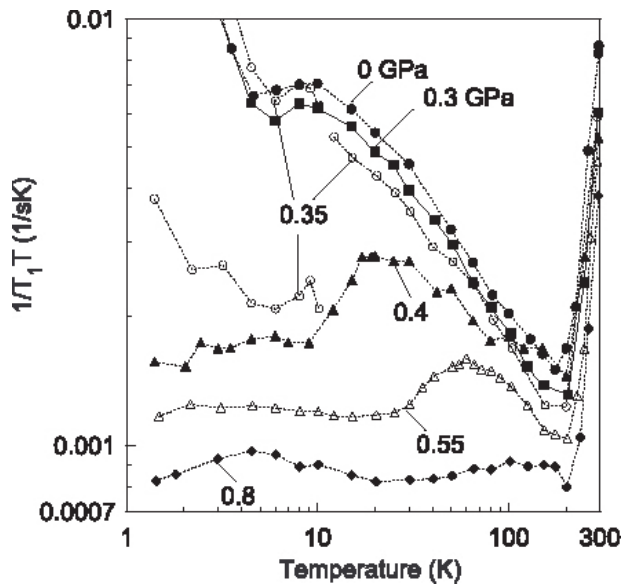}
	\caption{\label{Fig4}
	The $^1$H NMR spin-lattice relaxation rate divided temperature, $1/T_1T$, of $\kappa$-(ET)$_2$Cu$_2$(CN)$_3$ at various pressures.
	 }
	\end{figure}
	Figure 4 shows the spin-lattice relaxation rate divided by temperature, $1/T_1T$. 
	The enhancement of $1/T_1T$ above 200 K arises from the thermally activated vibration of the ethylene groups of the ET molecule.
	It is well known that this motional contribution to $1/T_1$ diminishes with the activation temperature dependence called Bloembergen-Purcell-Pound type \cite{Ref15}, as seen in other ET salts \cite{Ref16}. 
	Figure 4 indicates that the motional contribution to $1/T_1$ at a NMR frequency (156 MHz) is overwhelmed by the electronic contribution below 150 K.
	At ambient pressure, $1/T_1T$ increases on cooling with a shoulder around 10 K. 
	At 0.3 GPa, the temperature dependence of $1/T_1T$ falls into the ambient-pressure data, although the values are slightly depressed in the whole temperature range. 
	A drastic change occurs at 0.4 GPa.
	$1/T_1T$ shows a sudden drop around 15 K and remains constant below 9 K. 
	From the resistance at 0.4 GPa, we conclude that the drop of $1/T_1T$ is a manifestation of the Mott transition and the temperature-independent $1/T_1T$ points to the Korringa's relation expected in Fermi liquid state. 
	No sign of superconductivity is due to the applied magnetic field of 3.6 T exceeding the upper critical field. 
	At a higher pressure of 0.55 GPa, $1/T_1T$ forms a broad peak around 60 K and follows the Korringa's law below 25 K. 
	Then $1/T_1T$ becomes nearly temperature independent under 0.8 GPa. 
	We also find that the inflection point of the resistance is close to the temperature giving the peak in $1/T_1T$ at 0.4 GPa and 0.55 GPa, as shown in Fig.1. 
	These broad peak structures in $1/T_1T$ was also observed in other members, $\kappa$-(ET)$_2$Cu(NCS)$_2$ and $\kappa$-(ET)$_2$Cu[N(CN)$_2$]Br  \cite{Ref17}.
	This correspondence means that the spin fluctuations are suppressed over the wave vector ${\bf q}$, as the electrons are delocalized. 
	Moreover, the uppermost temperatures, below which $1/T_1T$ = const. is satisfied, agrees with the Fermi liquid-bad metal boundary seen in the resistance measurements, as plotted in Fig. 1. 

	At 0.35 GPa, which is close to the critical pressure of Mott transition, two components of $1/T_1T$ with the comparable fractions emerge only below 10 K. 
	One component behaves similarly to the ambient-pressure data, while the other is close to that at 0.4 GPa. 
	This suggests the phase segregation of the insulating and metallic phases at low temperatures, reflecting the first-order Mott transition or inhomogeneity in the internal pressure and/or the sample. 
	It is reminiscent of the metal-insulator phase segregation observed by the $^{13}$C NMR measurements in $\kappa$-(deutrated-ET)$_2$Cu[N(CN)$_2$]Br located just on the Mott boundary \cite{Ref18}. 

	\begin{figure}
	\includegraphics[scale=0.38]{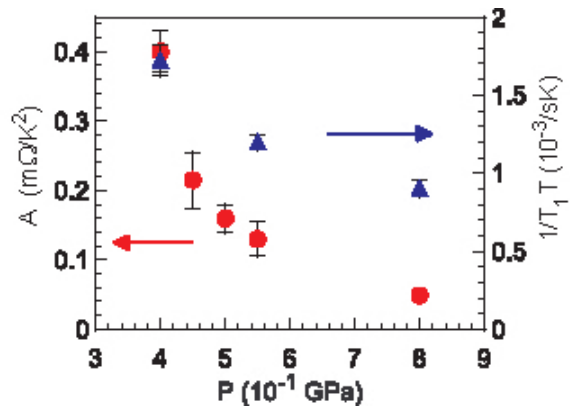}
	\caption{\label{Fig5}
	The pressure dependence of $A$ and $1/T_1T$ at low-temperature region. Closed circles and triangles represent the $A$ and $(T_1T)^{-1}$ values, respectively. 
	}
\end{figure}
	The pressure dependence of $A$ and $1/T_1T$ are shown in Fig.5. 
	We believe that these quantities are of fundamental importance in describing the Mott transition but still not fully explained theoretically. 
	If a simple Fermi-liquid regime is assumed, $A$ measures the electron-electron scattering rate and is proportional to square of the density of states at the Fermi level, $D(E_F)^2$, while $1/T_1T$ is proportional to both $D(E_F)^2$ and the so-called NMR enhancement factor $K(\alpha)$ that reflects spin fluctuations. 
	Both $A$ and $1/T_1T$ decrease with increasing pressure, which suggests that $D(E_F)$ is suppressed as the system goes away from the Mott boundary. 
	The similar pressure dependence of $A$ was observed in $\kappa$-(ET)$_2$Cu[N(CN)$_2$]Cl \cite{Ref11}, and the $1/T_1T$ values of $\sim$0.0011 s$^{-1}$K$^{-1}$ in $\kappa$-(ET)$_2$Cu[N(CN)$_2$]Br \cite{Ref19} and $\sim$0.0009 s$^{-1}$K$^{-1}$  in $\kappa$-(ET)$_2$Cu(NCS)$_2$ \cite{Ref20} are reasonably situated on the pressure dependence of $1/T_1T$ in Fig.\ref{Fig5}. 
	The pressure dependence of $1/T_1T$ is more moderate than that of $A$.
	This is tentatively attributed to the pressure dependence of $K(\alpha)$; the AFM fluctuations are enhanced by pressure.
	The scenario is not unlikely, if the nearly isotropic triangular lattice is deformed by pressure, but may oversimplify the situation because the spin fluctuations should also affect the $A$ term.
	Further theoretical studies are required to understand the transport and magnetic properties under the large spin and charge fluctuations. 

	To figure out the effect of the spin frustration on the Mott transition, it is useful to compare the phase diagram of $\kappa$-(ET)$_2$Cu$_2$(CN)$_3$ with those of the typical Mott insulators such as $\kappa$-(ET)$_2$Cu[N(CN)$_2$]Cl \cite{Ref10, Ref11, Ref12} and V$_2$O$_3$ \cite{Ref21}, in which the insulating phase shows the AFM transition at a finite temperature. 
	The Mott transition lines of $\kappa$-(ET)$_2$Cu[N(CN)$_2$]Cl and V$_2$O$_3$ have a positive gradient ($dT/dP$ $>$ 0) at high temperatures, but it turns negative at low temperatures where the AFM ordering occurs in the insulating phase. 
	On the other hand, for $\kappa$-(ET)$_2$Cu$_2$(CN)$_3$ the gradient, $dT/dP$, keeps positive in the whole temperature region investigated. 
	We repeated the resistance measurements with finely tuning the pressure around the critical pressure and confirmed that the insulator-metal-insulator re-entrant transition is absent as observed in $\kappa$-(ET)$_2$Cu[N(CN)$_2$]Cl. 
	The difference can be explained in terms of the Clasius-Clapeyron relation, $dT/dP$ = $\Delta V/\Delta S$ \cite{Ref10}. 
	Here  $\Delta V$ = $V_{ins}$ - $V_{metal}$ and $\Delta S$ = $S_{ins}$ - $S_{metal}$ are the volume and entropy changes on the Mott transition. 
	Since the metallic phase appears under pressure, $V_{metal}$ should be smaller than $V_{ins}$; $\Delta V$ $>$ 0. 
	The entropy at low temperatures can be almost attributed to the spin degrees of freedom. 
	The large spin entropy of paramagnetic Mott insulator with the magnetic order at a finite temperature rapidly dies away with decreasing temperature, because of the AFM fluctuations and order, while the entropy of the metal is proportional to temperature. 
	Hence $\Delta S$ can change from positive to negative for $\kappa$-(ET)$_2$Cu[N(CN)$_2$]Cl and V$_2$O$_3$ with AFM ordered phase. 
	In case of $\kappa$-(ET)$_2$Cu$_2$(CN)$_3$, however, $\Delta S$ may well remain positive even at low temperatures owing to the suppression of magnetic order by the spin frustration, leading to positive $dT/dP$. 
	Thus the spin degrees of freedom have an appreciable influence on the Mott transition at low temperatures. 
	The vertical slope of $dT/dP$ in the low-temperature limit is consistent with the thermodynamics of Mott transition \cite{Watanabe}. 

	In conclusion, we have investigated the $P-T$ phase diagram of the triangular-lattice Mott insulator $\kappa$-(ET)$_2$Cu$_2$(CN)$_3$. 
	We found that magnetic order is absent in the insulating phase up to the pressure of the insulator-metal transition boundary. 
	The Mott transition was seen to occur between spin liquid state and Fermi liquid state. 
	The low-temperature feature in the constructed phase diagram points to the important role of the spin frustration in the Mott transition. 
	The superconductivity neighboring the spin liquid in this material is intriguing as a future issue and recently attracting theoretical attention \cite{Kondo, Liu}.

	We thank F. Kagawa, T. Itou, M. Maesato, J. Shinagawa, M. Imada, N. Nagaosa, H. Fukuyama and T. Moriya for stimulating discussions. 
	This work was partially supported by MEXT KAKENHI on Priority Area of Molecular Conductors (No.15073204), JSPS KAKENHI (No.15104006) and COE project. 

\end{document}